\documentclass[aps,twocolumn,groupedaddress,showpacs]{revtex4}

\usepackage{graphicx}%
\usepackage{dcolumn}
\usepackage{amsmath}

\begin{document}
\bibliographystyle{apsrev}


\title{Quantum annealing of the random-field Ising model
by transverse ferromagnetic interactions}


\author{Sei Suzuki}
\email[]{sei@stat.phys.titech.ac.jp}
\author{Hidetoshi Nishimori}
\affiliation{Department of Physics, Tokyo Institute of Technology,
Oh-okayama, Meguro, Tokyo 152-8551, Japan}
\author{Masuo Suzuki}
\affiliation{Department of Applied Physics, 
Tokyo University of Science, Kagurazaka, Shinjyuku, Tokyo 162-8601,
Japan}


\date{\today}

\begin{abstract}
We introduce transverse ferromagnetic interactions, in addition to a simple
transverse field, to quantum annealing of the random-field Ising model to
accelerate convergence toward the target ground state.
The conventional approach using only the transverse-field term is known
to be plagued by slow convergence when the true ground state
has strong ferromagnetic characteristics for the random-field Ising model.
The transverse ferromagnetic interactions are shown
to improve the performance significantly in such cases.
This conclusion is drawn from the analyses of the energy eigenvalues of
instantaneous stationary states as well as by the very fast algorithm of
Bethe-type mean-field annealing adopted to quantum systems.
The present study highlights the importance of a flexible choice
of the type of quantum fluctuations to achieve the best possible
performance in quantum annealing.
The existence of such flexibility is an outstanding advantage of
quantum annealing over simulated annealing.

\end{abstract}
\pacs{05.50.+q,75.50.Lk,75.10.Jm,02.60.Pn}

\maketitle

\section{Introduction}

Combinatorial optimization is one of the central problems in computer
science \cite{bib:Garey}.
A celebrated example is the traveling salesman problem, and many instances
as well as this problem can be represented in
terms of the Ising model, often with disorder or
frustration \cite{bib:Hopfield}. 
In such cases, optimization corresponds to the search of the ground state
of the disordered/frustrated Ising model,
a highly non-trivial task also from the viewpoint of physics.

Simulated annealing is a generic prescription suited for such purposes
\cite{bib:Kirkpatrick}.
One introduces an artificial temperature variable to control the average
energy through thermal fluctuations.
By decreasing the temperature gradually from a very high initial value,
one hopes to eventually reach the ground state in the zero-temperature limit.

This prescription of simulated annealing intrinsically restricts the control
parameter only to the temperature.
Also, the actual implementation of simulated annealing is exclusively
realized by classical Monte Carlo simulations with time-dependent temperatures.

In contrast, quantum annealing makes use of quantum fluctuations to
control the behavior of the relevant system
\cite{bib:Kadowaki,bib:QABook,bib:SantoroReview,bib:Finnila}.
Fluctuations of quantum nature are introduced artificially to the classical
optimization problem represented by a disordered/frustrated Ising model.
One initially sets the coefficient of the quantum term very large
so that the system searches very wide regions of the phase space
for the optimal state.
A gradual decrease of the coefficient of the quantum term yields
an adiabatic evolution of the ground state starting
from a trivial initial ground state, and leads to the non-trivial
final ground state, which is the solution of the original optimization
problem,.
We have some degrees of freedom in the choice of
a type of quantum fluctuation.
In addition, the implementation of quantum annealing can be made in a
variety of ways including an exact numerical integration of the
Schr\"odinger equation (for small systems), quantum Monte Carlo
simulations \cite{bib:KadowakiPhD,bib:Martonak},
and Green's function Monte Carlo method \cite{bib:StellaGFMC}.

Most studies of quantum annealing so far have put an emphasis on how fast
quantum annealing performs, relatively to simulated annealing, with
a single type of the quantum term in most cases, 
namely the transverse-field term
added to the classical Ising model representing the original optimization problem.
Evidence has been accumulating to establish generic superiority
of quantum annealing
\cite{bib:Kadowaki,bib:Santoro,bib:MartonakTSP}. 

In the present paper we focus our attention on a different aspect of quantum
annealing, flexibility in the choice of quantum terms.
In particular, we investigate the random-field Ising model, for which
quantum annealing with the conventional type of quantum term
has been shown not to perform efficiently under certain circumstances
\cite{bib:Sarjala}.
We introduce additional terms of quantum nature, transverse interactions
of a ferromagnetic type, by taking advantage of the flexibility in quantum annealing.
The result shows a significant improvement, from which we conclude
that the reduced efficiency of quantum annealing in comparison with
simulated annealing reported in some cases may not necessarily reflect
intrinsic limitations of quantum annealing.
Rather, we expect that an appropriate choice of quantum terms
would often accelerate  significantly the convergence of quantum annealing to the ground state.
This feature of flexibility is an outstanding advantage of quantum annealing.

We organize this paper as follows.
In Sec. \ref{sec:QA}, we first explain the basic ideas of conventional quantum 
annealing and quantum annealing using transverse ferromagnetic interactions.
We then analyze
the instantaneous ground state and the first excited state
by numerical diagonalization.
Advantages of transverse ferromagnetic interactions as a driving force
of quantum annealing are pointed out.
In Sec. \ref{sec:perform}, 
quantum annealing by transverse ferromagnetic interactions
is carried out for systems of large size using the Bethe approximation
as an algorithm for practical implementation.
We then present results of simulations
and compare residual errors of conventional quantum annealing,
quantum annealing by transverse ferromagnetic interactions,
and simulated annealing.
Section \ref{sec:conclusion} is devoted to conclusion.

\section{Transverse ferromagnetic interactions}
\label{sec:QA}
\subsection{Conventional approach}

In the present paper
we consider the problem of ground-state search of the random-field Ising model.
The Hamiltonian
\begin{equation}
 \mathcal{H}_{\rm pot}^{\rm RFIM} 
= - J \sum_{< i j >} \sigma^z_i \sigma^z_j
- \sum_i h_i \sigma^z_i ,
\label{eq:HRFIM1}
\end{equation}
where $\sigma^z_i$ is the $z$ component of the Pauli matrix,
is regarded as the classical potential term in the context of quantum annealing.
For this system, it is known that the conventional approach to quantum annealing using
the transverse field as described below
does not perform impressively
when the interaction term in Eq. (\ref{eq:HRFIM1}) is dominant
relatively to the random-field term \cite{bib:Sarjala}.
We choose the two-dimensional square lattice
and the random field is assumed to be either $+1$ or $-1$
with equal probability.

In the conventional approach, one adds a term of transverse field
to Eq. (\ref{eq:HRFIM1})
\begin{equation}
 \mathcal{H}_{\rm kin}^{\rm TF} = - \sum_i \sigma^x_i ,
\label{eq:TF_kin}
\end{equation}
which may also be regarded as the quantum kinetic energy,
to induce quantum transitions between classical states.
The strength of the quantum kinetic energy is controlled in
the present paper by a linear function of time,
\begin{equation}
 \mathcal{H}(t) = \left(1 - \frac{t}{\tau}\right)
  \mathcal{H}_{\rm kin}^{\rm TF}
 + \frac{t}{\tau}\mathcal{H}_{\rm pot}^{\rm RFIM}
\label{eq:HofT}
\end{equation}
with $\tau$ being the time assigned for annealing.
The initial state is chosen to be the trivial ground state
of the kinetic term
$
 |\Psi(0)\rangle = \bigotimes_i [(|\uparrow\rangle_i + 
|\downarrow\rangle_i)/\sqrt{2}]
$,
where $|\uparrow\rangle_i$ and $|\downarrow\rangle_i$ are
the eigenstates of $\sigma^z_i$ with the eigenvalues $1$ and $-1$,
respectively.
It is expected intuitively that the system evolves adiabatically
following the instantaneous ground state if $\tau$ is very large,
reaching finally the non-trivial ground state of the potential
term $\mathcal{H}_{\rm pot}^{\rm RFIM}$ at $t=\tau$
\cite{bib:FarhiCondMat}.

The quantitative criterion for such an adiabatic evolution is
\begin{equation}
 \tau \gg \tau_{\rm c} \equiv  \frac{1}{\varepsilon_{\rm min}^2}
 \left|\left\langle\Psi_1\left(\frac{t}{\tau}\right)\right| 
 \frac{d \mathcal{H}(t)}{d (t/\tau)}\left|\Psi_0\left(\frac{t}{\tau}\right)\right\rangle
  \right|_{\rm max} ,
\label{eq:TauC}
\end{equation}
where $\left|\langle\Psi_1(t/\tau)|A| \Psi_0(t/\tau)\rangle\right|_{\rm max}$ 
denotes the maximum absolute value of
the matrix element of $A$ between the instantaneous ground state $\Psi_0$
and the first excited state  $\Psi_1$. 
The numerator $\varepsilon_{\rm min}$ is the minimum energy gap between the instantaneous
ground state and the first excited state.
The residual error, the difference between the obtained approximate
energy at $t=\tau$ and the true ground state energy of $\mathcal{H}_{\rm pot}^{\rm RFIM}$,
is of $\mathcal{O}((\tau_{\rm c}/{\tau})^2)$
if the condition (\ref{eq:TauC}) is satisfied \cite{bib:Sei}.

It is therefore important to find an improved quantum kinetic term
to enhance the gap $\varepsilon_{\rm min}$ and suppress the transition
matrix element $\left|\langle\Psi_1(t/\tau)|\cdot | \Psi_0(t/\tau)\rangle\right|_{\rm max}$.
Notice that the potential term $\mathcal{H}_{\rm pot}^{\rm RFIM}$
is given and fixed, and our degree
of freedom lies in the choice of the quantum kinetic term.
This flexibility is an outstanding feature of quantum annealing,
which has not necessarily been fully exploited in existing studies
using almost always the transverse-field term of Eq. (\ref{eq:TF_kin}).

\subsection{Transverse ferromagnetic interactions}

We now try the following operator as the quantum kinetic term
in place of $\mathcal{H}_{\rm kin}^{\rm TF}$ of Eq. (\ref{eq:TF_kin}),
\begin{equation}
 \mathcal{H}_{\rm kin}^{\rm FI} = -\sum_i \sigma^x_i -
 \sum_{< i j >} \sigma^x_i \sigma^x_j .
 \label{eq:HFI1}
\end{equation}
The time evolution of the system is governed by the time-dependent
Hamiltonian similar to Eq. (\ref{eq:HofT}),
\begin{equation}
 \mathcal{H}(t) = \left(1 - \frac{t}{\tau}\right)
  \mathcal{H}_{\rm kin}^{\rm FI}
 + \frac{t}{\tau}\mathcal{H}_{\rm pot}^{\rm RFIM}.
\label{eq:HofT2}
\end{equation}
In Eq. (\ref{eq:HFI1}) the sum for interactions runs over the same pairs
of sites as in the potential term (\ref{eq:HRFIM1}).
The ground state of Eq. (\ref{eq:HFI1}) is the trivial ferromagnetic state
in the $x$ direction as in the conventional case.
The overall symmetry (by the operation $\sigma_i^x\to -\sigma_i^x$) is
broken by the transverse field term
as is necessary to guarantee the uniqueness of the ground state.
For simplicity we fix the coefficients to unity in  Eq. (\ref{eq:HFI1})
and do not try to optimize the ratio of the coefficients
of the transverse field and interaction terms.

\begin{figure}
\begin{center}
\includegraphics[width=8.5cm]{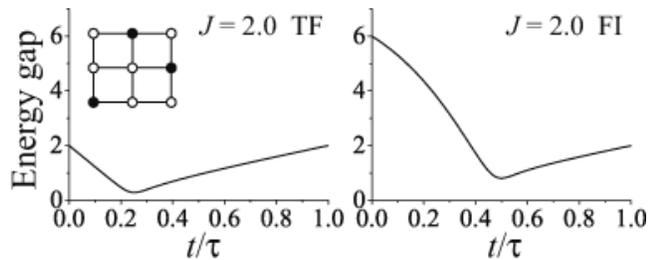}
\caption{Instantaneous energy gaps between the ground state and the first
excited state of the time-dependent Hamiltonians for systems with $N=9$ spins.
The coupling constant in $\mathcal{H}_{\rm pot}^{\rm RFIM}$ is
set at $J=2.0$.
`TF' and `FI' in the figures correspond to the kinetic Hamiltonians
$\mathcal{H}_{\rm kin}^{\rm TF}$ and $\mathcal{H}_{\rm kin}^{\rm FI}$,
respectively.
The configuration of random fields is depicted in the left panel, where
open circles indicate $h_i = +1$ and filled ones $h_i = -1$.
}
\label{fig:09J20RFGap}
\end{center}
\end{figure}

\begin{figure}
\begin{center}
\includegraphics[width=8.5cm]{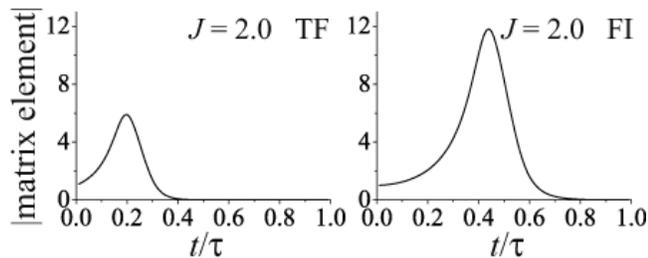}
\caption{Absolute values of the matrix element,
$|\langle\Psi_1(t/\tau)|d\mathcal{H}(t)/d(t/\tau)|
\Psi_0(t/\tau)\rangle|$, for the system of Fig. \ref{fig:09J20RFGap}.
}
\label{fig:09J20RFElm}
\end{center}
\end{figure}

As a preliminary analysis, we have evaluated the characteristic time $\tau_{\rm c}$ of
Eq. (\ref{eq:TauC}) by direct diagonalization of $\mathcal{H}(t)$ for small systems.
Figure \ref{fig:09J20RFGap} shows the energy gap between the instantaneous
ground state and the first excited state for the system with nine spins ($N=9$)
under the configuration of random fields indicated on the left panel.
The interaction constant is $J=2.0$.
It is observed that the minimum value of the gap $\varepsilon_{\rm min}$ is larger
for the case of transverse interactions (FI) on the right panel than
for the transverse field (TF) case on the left panel:
$\varepsilon_{\rm min}^{\rm FI}/\varepsilon_{\rm min}^{\rm TF}=2.89$.
The absolute value of the matrix element appearing in Eq. (\ref{eq:TauC})
is depicted in Fig. \ref{fig:09J20RFElm}.
This quantity is also larger for FI than for TF, the ratio being 2.01.
Thus the ratio of $\tau_{\rm c}$ is $\tau_{\rm c}^{\rm FI}/\tau_{\rm c}^{\rm TF}=
2.01/2.89^2=0.24$.
We therefore conclude that quantum fluctuations by transverse ferromagnetic
interactions lead to much shorter characteristic time than the
conventional counterpart for the present system.
A similar tendency was observed for a larger system with $N=20$ and $J=2.0$.

\begin{figure}
\begin{center}
\includegraphics[width=8.5cm]{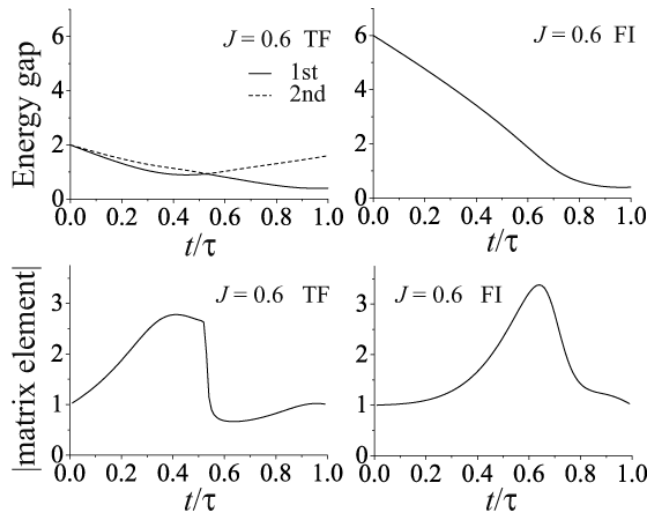}
\caption{Instantaneous energy gaps (upper panel)
and absolute values of the matrix element (lower panel)
for the random field configuration of Fig. \ref{fig:09J20RFGap}
with $J=0.6$.
The energy gap between the second excited state and the ground state
is also plotted in the panel for TF.
The abrupt change of the matrix element in the TF panel is due to level
crossing of the first and second excited states.
}
\label{fig:09J06RF}
\end{center}
\end{figure}

The situation changes for $J=0.6$ as is seen in Fig. \ref{fig:09J06RF}.
The ratio of characteristic times is now
$\tau_{\rm c}^{\rm FI}/\tau_{\rm c}^{\rm TF}=
1.21$, suggesting a slightly slower convergence under the transverse
ferromagnetic interactions.
Nevertheless this deterioration by 21\% may not be quite significant in
comparison with the much larger gain of 76\% ($=1-0.24$) for $J=2.0$.

In the next section we shall see the consequences of these observations
from a different point of view.

\section{Mean-field annealing by the Bethe approximation}
\label{sec:perform}

Implementation of quantum annealing for large systems is often
carried out by quantum Monte Carlo simulations
based on the Suzuki-Trotter transformation \cite{bib:SuzukiTrotter}.
In the present paper we instead make use of the method of mean-field annealing
with the Bethe-type approximation \cite{bib:Tanaka}.
The basic idea is to solve a set of equations for local magnetization iteratively.
There are several reasons for the choice of this method.
First, our main objective is to study the {\em relative} effectiveness of
transverse ferromagnetic interactions and transverse field as the quantum
kinetic energy.
The present approximation is likely to affect both approaches to a similar
degree, and therefore the relative performance is expected to be largely
unchanged by the introduction of approximation.
Second, the mean-field annealing is much faster than quantum Monte Carlo.
The former may therefore be suitable for practical purposes
under appropriate circumstances.
Third, we can follow the ground state directly in the mean-field annealing
without introducing small but finite temperatures as in quantum Monte Carlo.
Lastly, the implementation of quantum Monte Carlo involving transverse
ferromagnetic interactions is a little more complicated than the
case of simple transverse field only, whereas mean-field annealing
can be formulated easily for both cases.

\subsection{Bethe approximation}

\begin{figure}
\begin{center}
\includegraphics[width=4cm]{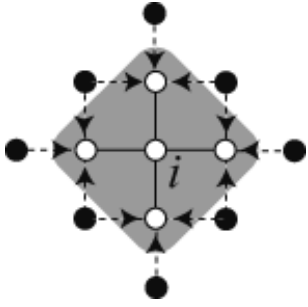}
\caption{In the Bethe approximation one focuses on a site $i$ and
its immediate neighbors.  All other site variables are
replaced with their averages.
}
\label{fig:Bethe}
\end{center}
\end{figure}

To achieve the best possible results within mean-field-type methods, we
employ the Bethe approximation in place of a simple single-body
mean-field theory.
One focuses on a site $i$ and approximates the Hamiltonian
involving $i$ by the cluster Hamiltonian (see Fig. \ref{fig:Bethe})
\begin{equation}
 \mathcal{H}^{(i)}(t) = \left(1 - \frac{t}{\tau}\right)
  \mathcal{H}_{\rm kin}^{(i)}+ \frac{t}{\tau}\mathcal{H}_{\rm pot}^{(i)}
\end{equation}
\begin{eqnarray}
 \mathcal{H}_{\rm pot}^{(i)}&=&-J\sigma^z_i \sum_{j\in\mathcal{S}(i)} \sigma^z_j 
 - h_i \sigma^z_i \nonumber\\
 &&-\sum_{j\in\mathcal{S}(i)} \left( h_j 
 +J\sum_{k\in\mathcal{S}(j)\setminus i} m_k^z \right) \sigma_j^z \\
 \mathcal{H}_{\rm kin}^{(i)}&=&-\sigma^x_i \sum_{j\in\mathcal{S}(i)}  \sigma^x_j 
 -  \sigma^x_i \nonumber\\
 &&-\sum_{j\in\mathcal{S}(i)}\left( 1
 +\sum_{k\in\mathcal{S}(j)\setminus i} m_k^x\right) \sigma_j^x ,
\end{eqnarray}
where $j$ is a neighboring site of $i$ and $k$ is a neighbor of $j$
excepting $i$.
For this Hamiltonian one calculates the ground-state expectation values of
$\sigma_i^z, \sigma_j^z, \sigma_i^x$ and $\sigma_j^x$, giving
$m_i^z, m_j^z, m_i^x$ and $m_j^x$, respectively.
The same process is repeated by shifting the center site $i$ of the cluster.
After scanning the whole system, one updates the time variable.
The same idea applies to the conventional quantum annealing, in which
one drops the transverse interaction term.

For comparison the same Bethe-type mean-field annealing was tested
for simulated annealing using thermal fluctuations.
In this case the kinetic term $\mathcal{H}_{\rm kin}^{(i)}$ is
dropped and temperature is introduced to calculate the thermal
expectation value $m_i^z$ of $\sigma_i^z$.
The same linear annealing schedule $T=T_0\cdot (1-t/\tau)$ was
used with a sufficiently high initial temperature $T_0$.

\subsection{Results}
We have applied the method above described to the random-field Ising
model with the system size $100\times 100$ 
in two dimensions.
The residual energy was calculated as the difference between the obtained
approximate energy and the true ground state energy estimated by a
well-established algorithm \cite{bib:Alava}.

\begin{figure}[tbp]
\includegraphics[width=8cm]{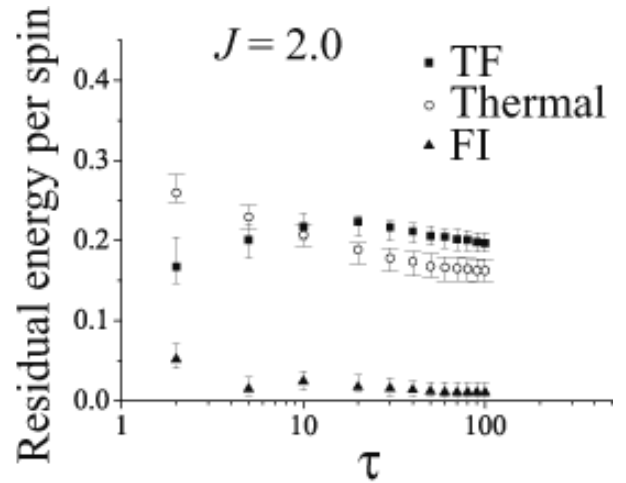}
\caption{Residual energy for the random field Ising model with $J=2.0$
averaged over 80 configurations of random fields.
Filled squares and triangles indicate the results of
quantum annealing by transverse interactions (TF) and by
transverse field (FI), respectively.
Open circles are for simulated annealing.
}
\label{fig:MFBJ20}
\end{figure}

\begin{figure}[tbp]
\includegraphics[width=7cm]{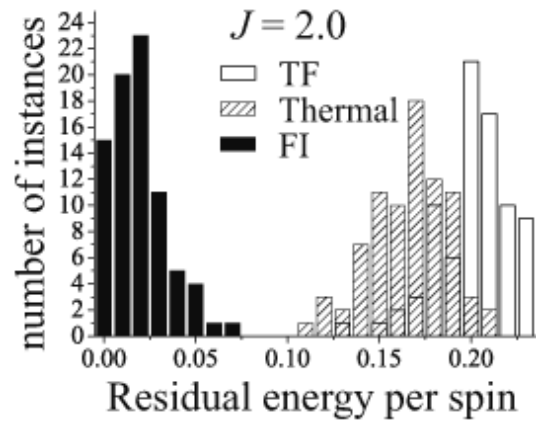}
\caption{Histogram of the residual energy at $\tau=100$. 
The parameters are the same as in  Fig. \ref{fig:MFBJ20}.
}
\label{fig:MFBJ20HG}
\end{figure}

Figure \ref{fig:MFBJ20} shows the result for $J=2.0$ averaged over
80 samples, and Fig. \ref{fig:MFBJ20HG} is the histogram of
the residual energy at $\tau=100$.
It is clearly seen that quantum annealing by transverse ferromagnetic
interactions is far more superior to the other methods.
Simulated annealing and the conventional transverse-field quantum annealing
perform almost similarly, the former being slightly better probably
owing to the difference in the effects of our approximation to thermal
and quantum annealing.

\begin{figure}[tbp]
\includegraphics[width=8cm]{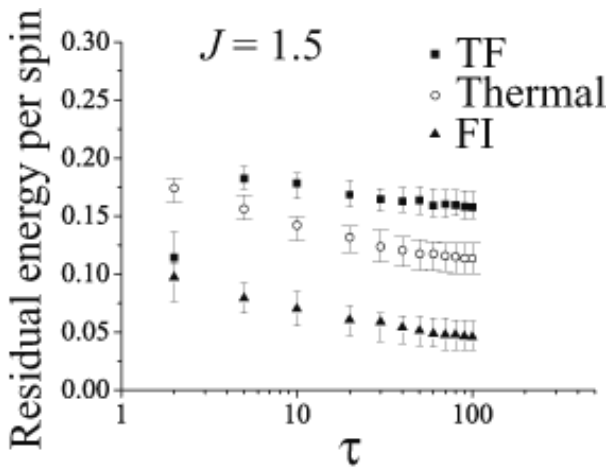}
\caption{Residual energy for  $J=1.5$.
Other conditions are the same as in Fig. \ref{fig:MFBJ20}.
}
\label{fig:MFBJ15}
\end{figure}

\begin{figure}[tbp]
\includegraphics[width=8cm]{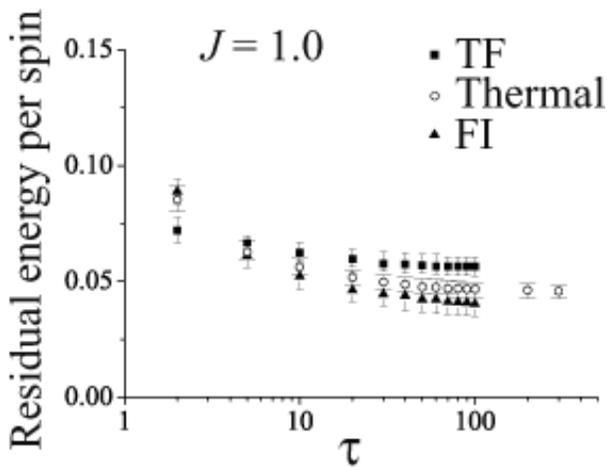}
\caption{Residual energy for $J=1.0$.
}
\label{fig:MFBJ10}
\end{figure}

\begin{figure}[tbp]
\includegraphics[width=8cm]{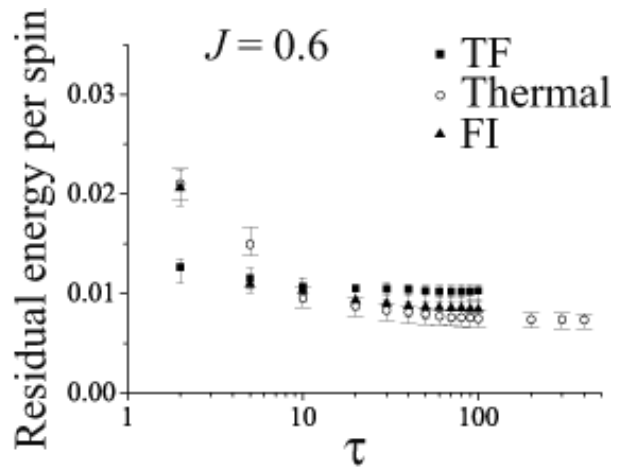}
\caption{Residual energy for $J=0.6$.
}
\label{fig:MFBJ06}
\end{figure}

A similar situation is found for $J=1.5$ in Fig. \ref{fig:MFBJ15}.
The difference between various methods diminishes as $J$ decreases further,
see Figs. \ref{fig:MFBJ10} and \ref{fig:MFBJ06}

Our method works efficiently for large $J$,
in which the true ground state is either exactly or close to the
ferromagnetic state.
Introduction of transverse ferromagnetic interactions does not
necessarily improve the result for the cases of disordered ground states.
This observation is to be compared with the finding of Sarjala 
\textit{et al}. \cite{bib:Sarjala}
who showed by quantum Monte Carlo simulations that quantum annealing
by the conventional method is less efficient than
simulated annealing when the ground state is strongly ferromagnetic.
We may conclude that their consequence does not necessarily reflect
intrinsic features of quantum annealing.
Sophisticated implementations of quantum annealing can significantly
improve the performance.
The same may apply to the problem of 3-SAT, in which quantum annealing
was observed to give less efficient results than simulated annealing 
\cite{bib:Battaglia}.

\section{Conclusion}
\label{sec:conclusion}
We have discussed the effect of the introduction of 
transverse ferromagnetic interactions to
quantum annealing of the random field Ising model.
The exact diagonalization study on small systems 
revealed that transverse ferromagnetic interactions shorten
the characteristic time, $\tau_{\rm c}$, from the value
given by the conventional transverse field when the interaction constant
$J$ is large.
Then we compared residual energies after quantum annealing
by transverse ferromagnetic interactions with those by
conventional quantum annealing using transverse field and 
simulated annealing for large systems.
In order to carry out quantum and thermal annealing in larger systems,
we employed the method of mean-field annealing based upon the Bethe-type
approximation.
The results of mean-field annealing showed that quantum annealing
by transverse ferromagnetic interactions is far more efficient than
the other two schemes 
for large $J$.
This implies that the previously reported fact on the 
reduced efficiency
of conventional quantum annealing for the ferromagnetic ground state
is not an intrinsic feature of quantum annealing, but a better efficiency
than simulated annealing can be obtained by exploiting an appropriate
quantum effect. 
It is important to make use of the room of choice of quantum fluctuations
to extract the best performance in quantum annealing.
Such flexibility is an outstanding character of
quantum annealing that does not exist in simulated annealing.
It may be also interesting to take into account many-body interactions
of the form $-\sigma_{i_1}^x\sigma_{i_2}^x\cdots\sigma_{i_n}$
in the mean-field quantum annealing.

\begin{acknowledgments}
 The present study was partially supported by CREST, JST. and
by a Grant-in-Aid for Scientific Research on Priority Areas.
\end{acknowledgments}


\end{document}